\documentclass[12pt,twoside]{article}
\usepackage{verbatim}
\usepackage{epsfig}
\usepackage{epstopdf}
\input{epsf.sty}
\usepackage{epsf}
\usepackage{amssymb}
\usepackage{amsfonts}
\usepackage{amsmath}
\usepackage{pifont}
\usepackage[spanish,english]{babel}
\usepackage{graphicx}
\usepackage{subfigure}

\begin{document}

\title{Scalar Dark Matter with CERN-LEP data and $Z'$  search at the LHC in an $U(1)'$ Model}
\author{R. Mart\'{\i}nez$\thanks{%
e-mail: remartinezm@unal.edu.co}$, J. Nisperuza$\thanks{e-mail: jlnisperu@unal.edu.co}$, F. Ochoa$\thanks{%
e-mail: faochoap@unal.edu.co}$, J. P. Rubio$\thanks{e-mail: jprubioo@unal.edu.co}$ \and Departamento de F\'{\i}sica, Universidad
Nacional de Colombia, \\ Ciudad Universitaria, Bogot\'{a} D.C.}
 
\maketitle

\begin{abstract}
In the framework of an nonuniversal $U(1)'$ extension of the standard model, we propose an scalar candidate for cold dark matter which exhibits interactions with ordinary matter through Higgs and gauge bosons. Using limits from low energy observables, we find constraints on the new physics parameters of the model associated to the extra abelian symmetry, in particular, the mass of the additional neutral gauge boson $Z'$ and the new gauge coupling constant. We found that for the lower experimental limit $M_{Z'}=3$ TeV constrained by direct research at LHC, the ratio between the $U(1)'$ and $SU(2)_L$ gauge coupling constants is around $0.4$.  Taking into account this limit and the observable relic density of the Universe, we search for new constraints on the mass of the dark matter particle of the model. We found that for a higgsphobic model, the mass of the scalar dark matter must be $M_{\sigma}=70$ GeV. We also found different kinematical threshold and resonances that enhance the dispersion of dark matter into ordinary matter for different regions of the space of parameters of the model, obtaining masses as low as $1.3$ GeV and as large as $125$ GeV with not allowed intermediate regions due to resonances.  
\end{abstract}

\section{Introduction}

Although the standard model (SM) \cite{SM} is the simplest model that successfully explains most of the phenomena and experimental observations in particle physics, there are still some unexplained observations and theoretical issues that leaves unanswered. In particular, the astrophysical evidences of electrically neutral, non-baryonic and weakly interacting matter, i.e. dark matter (DM) can be naturally understood if the SM is extended. Since the evidences of DM are based only on its gravitational effects, 
its nature and microscopic properties remains unknown. Thus, in particle physics many extensions of the SM can be proposed with new particles that naturally fits the current astrophysical data and experimental constraints in direct DM detection. Among the different candidates for DM, extensions with stable and non-relativistic (cold) particles with masses between a few GeV and 1 TeV, and interacting through the weak nuclear force (WIMPs) are currently the best motivated models which provides a natural scenery for the cold dark matter (CDM). For example, supersymmetry provides a WIMP candidate through the lightest neutralino \cite{SUSY}. In universal extra dimension models, the lightest Kaluza-Klein partner is stable and a possible DM candidate \cite{KK}. Other models with WIMP candidates have been recently proposed, as for example branons in large extra dimensions \cite{LED}, T-odd particles in little Higgs models \cite{LitHiggs}, and excited states in warped extra dimensions \cite{WED}. Also, scalar candidates in scalar extensions of the SM through singlets and doublets have been widely considered \cite{scalar} in the literature, and more recently in 3-3-1 models \cite{331}. In particular, after the observation of the Higgs boson at the LHC \cite{atlas, cms}, the existence of scalar elementary particles in the Universe have been experimentally established. Thus, extensions of the SM with additional scalar sectors are interesting alternatives that may provide natural scenarios for CDM that match the experimental data.

On the other hand, models with extra $U(1)'$ symmetry are one of the most studied extensions of the SM, which implies many phenomenological and theoretical advantages including flavor physics \cite{flavor}, neutrino physics \cite{neutrino}, dark matter \cite{DM}, among other effects. A complete review of the above possibilities can be found in reference \cite{review}. In particular, family nonuniversal $U(1)'$ symmetry models have many well-established motivations. For example, they provide hints for solving the SM flavor puzzle, where even though all the fermions acquire masses at the same scale, $\upsilon =246$ GeV, experimentally they exhibit very different mass values. These models also imply a new $Z'$ neutral boson, which contains a large number of phenomenological consequences at low and high energies \cite{zprime-review}. In addition to the new neutral gauge boson $Z'$, an extended fermion spectrum is necessary in order to obtain an anomaly-free theory. Also, the new symmetry requires an extended scalar sector in order to (i) generate the breaking of the new Abelian symmetry and (ii) obtain heavy masses for the new $Z'$ gauge boson and the extra fermion content.

In this paper we consider an nonuniversal $U(1)'$ extension of the type from reference \cite{U1}, which exhibits an extended scalar sector with two scalar doublets and two singlets with nontrivial $U(1)'$ charges (labeled for this particular model as $U(1)_X$), where the lightest scalar singlet (denoted as $\sigma _0$) is taken as candidate for CDM. Some phenomenological consequences of this model have been studied in the above reference, with special emphasis in the neutral gauge and Yukawa sectors. Motivated by the problem of DM in astrophysics, further analyses of this model is continued. Among the most interesting features that the model exhibits, we mention the following related to DM: 

\begin{enumerate}

\item[(i)] The additional $Z'$ neutral boson produces low energy deviations through a $Z$-$Z'$ mixing angle. Restrictions to these deviations from the CERN-LEP and the SLAC collider impose restrictions on the mixing angle and the $U(1)_X$ coupling, as we will show. At the same time, since $\sigma _0$ scatter off via the $U(1)_X$ interaction, these parameters have effects on the DM relic density through its annihilation to neutral gauge boson $Z$. Thus, the model can impose constraints on astrophysical observables from low energy particle physics.

\item[(ii)] Although data from atomic parity violation additionaly imposes lower constraints on the mass of $Z'$ around $2$ TeV, the current limit on extra neutral gauge boson on direct research at LHC imposes a stronger limit above $3$ TeV, which have effects in the couplings of DM with the gauge boson $Z$.  

\item[(iv)] The nonuniversal feature of the quark sector with different $U(1)_X$ charges, could provide new signatures to search for DM presence 
in hadron colliders.        

\item[(v)] By implementing appropriate conserved discrete or continuos symmetries, we can both remove large flavor changing neutral current couplings and odd power couplings which ensure the stability of the DM candidate. At the same time, these symmetries preserve the couplings that allow the DM annihilation to SM particles required to fit the relic density observed in the Universe.
 
\end{enumerate}

Taking into account the most recent calculations of the Z pole observables in the SM, the new data from atomic parity violation, and the current limit on the extra neutral gauge boson research at LHC, we find constraints on some parameters associated to the $Z$-$Z'$ mixing angle and the $U(1)_X$ gauge coupling. Taking into account the above constraints, we obtain allowed regions for the DM relic density according to the last results from Planck experiment \cite{planck}. 

This paper is organized as follows. Sec. 2 is devoted to summarizing the spectrum of the model, the interactions with the light scalar singlet, and the basic condition for this scalar to be a CDM candidate. We also show the weak couplings, taking into account the small $Z$-$Z'$ mixing angle. In Sec. 3, we perform a global $\chi ^2$ fit at 95\% C.L to obtain constraints from the Z pole parameters. In Sec. 4, after summarizing the basic equations for the thermal relic density, we identify the fundamental interactions that determine the thermal average DM annihilation cross section. Using the constraints obtained previously from the Z pole analysis and using the software MicrOMEGAS \cite{micromegas}, we determine the relic density predicted by the model, and constraints on the DM mass for different ranges of the space of parameters according to the observed relic density by assuming that the scalar singlet provides the total composition of DM in the Universe.    

\section{The Model}

\subsection{Particle content}

The particle content of the model \cite{U1} is composed of ordinary SM particles and new exotic non-SM particles, as shown in Tables \ref{tab:SM-espectro} and \ref{tab:exotic-espectro}, respectively, where column $G_{sm}$ indicates the transformation rules under the SM gauge group $(SU(3)_c,SU(2)_L,U(1)_Y)$, column $U(1)_X$ contains the values of the new quantum number $X$, and in the column labeled {\it Feature}, we describe the type of field. Some properties of this spectrum are:

\begin{enumerate}

\item The $U(1)_X$ symmetry is nonuniversal only in the left-handed SM quark sector: the quark family $i=3$ has $X_3=1/3$ while families $i=1,2$ have $X_{1,2}=0$. To distinguish the top and bottom quarks from the other flavors, we use the following assignation for the phenomenological families:

\begin{equation}
U^{1,2,3}=(u,c,t), \hspace{1cm} D^{1,2,3}=(d,s,b), \hspace{1cm} e^{1,2,3}=(e,\mu ,\tau ). 
\label{quark-assignation}
\end{equation}

\item The SM leptons and scalar particles obtain nontrivial values for the new charge $X$.

\item In order to ensure cancellation of the gauge chiral anomalies, the model includes in the quark sector three extra singlets $T$ and $J^{n}$, where $n=1,2$. They are quasichiral, i.e. chiral under $U(1)_X$ and vector-like under $G_{sm}$.

\item In addition, to obtain a realistic model compatible with oscillation data, we include new neutrinos  $(\nu ^i_R)^c$ and $N ^i_R$ which may generate seesaw neutrino masses.

\item An additional scalar doublet $\phi _2$ identical to $\phi _1$ under  $G_{sm}$ but with different $U(1)_X$ charges is included in order to avoid massless charged fermions, and 
where the individual vacuum expectation values (VEVs)  are related to the electroweak VEV through the relation $\upsilon = \sqrt{\upsilon  _1 ^2+\upsilon _2^2}$.

\item An extra scalar singlet $\chi _0$ with VEV $\upsilon _{\chi}$ is required to produce the symmetry breaking of the $U(1)_X$ symmetry. We assume that it happens at a large scale $\upsilon _{\chi} \gg \upsilon$.

\item Another scalar singlet $\sigma _0$ is introduced. Since it is not essential for the symmetry breaking mechanisms, we may choose any value for its VEV. However, it is necessary that $\upsilon _{\sigma } = 0$ to ensure its stability. 

\item Finally, an extra neutral gauge boson $Z'_{\mu}$ is required to obtain a local $U(1)_X$ symmetry.

\end{enumerate}

With the above spectrum, we will determine all the interactions with the scalar particles of the model.

\subsection{Higgs potential}

The most general, renormalizable and $G_{sm} \times U(1)_X$ invariant potential is

\begin{eqnarray}
V&=&\mu _1^2 \left| \phi _1 \right|^2 +\mu _2^2 \left| \phi _2 \right|^2 + \mu _3^2 \left| \chi _0\right|^2 +\mu _4^2 \left| \sigma _0\right|^2+\mu _5^2\left( \chi _0^*\sigma _0 +h.c \right)  \nonumber \\
&+&f_1\left(\phi _2^{\dagger} \phi _1 \sigma _0+h.c.\right)+f_2\left(\phi _2^{\dagger} \phi _1 \chi _0+h.c.\right) \nonumber \\
&+& \lambda _1 \left| \phi _1 \right|^4+\lambda _2 \left| \phi _2 \right|^4+\lambda _3 \left| \chi _0\right|^4+\lambda _4 \left| \sigma _0\right|^4 \nonumber \\
&+& \left| \phi _1 \right|^2\left[ \lambda _6 \left| \chi _0\right| ^2+ \lambda '_6 \left| \sigma _0\right|^2+ \lambda ''_6 \left( \chi _0^*\sigma _0 +h.c. \right) \right] \nonumber \\
&+& \left| \phi _2 \right|^2\left[ \lambda _7 \left| \chi _0\right| ^2+ \lambda '_7 \left| \sigma _0\right|^2+ \lambda ''_7 \left( \chi _0^*\sigma _0+h.c. \right) \right] \nonumber \\
&+&\lambda _5 \left| \phi _1 \right|^2\left| \phi _2 \right|^2+\lambda '_5 \left| \phi _1^{\dagger} \phi _2 \right|^2 +\lambda _8  \left| \chi _0\right|^2\left| \sigma _0\right|^2+\lambda '_8  \left[\left(\chi _0^* \sigma _0\right)^2+h.c.\right].
\label{higgs-pot-1}
\end{eqnarray}
However, this potential is not suitable for scalar DM candidates. If we assume that the singlet $\sigma _0$ corresponds to the DM portion of the model, we must impose the following properties:

\begin{enumerate}

\item[(i)] Since $\sigma _0$ acquires nontrivial charge $U(1)_X$, it must be complex in order to obtain massive particles necessary for CDM. 

\item[(ii)] Terms involving odd powers of $\sigma _0$ (i.e., the $\mu _5$, $f_1$ and  $\lambda ''_{6,7}$ terms) induce decay of the DM, which spoils the prediction of the model for the DM relic density. Thus, we must impose either a global discrete or continuos symmetry. In particular, to reduce additional free parameters, we can eliminate the $\lambda '_8$ term by demanding that the potential respect the global symmetry

\begin{eqnarray}
\sigma _0 \rightarrow e^{i\theta }\sigma _0.
\label{global-symm}
\end{eqnarray}

\item[(iii)] In order to avoid the above symmetry to break spontaneously or new sources of decay, $\sigma _0$ must not generate VEV during the evolution of the Universe. Thus, we demand $\upsilon _{\sigma} = 0$. 

\end{enumerate}
With the above conditions, the potential is the same as in (\ref{higgs-pot-1}) but with $\mu _5=f_1=\lambda ''_{6,7}=\lambda '_8=0$. When we apply the minimum conditions $\partial \langle V \rangle/\partial \upsilon _i$ for each scalar VEV $\upsilon _i = \upsilon _{1,2,\chi }$, the following relations are obtained:

\begin{eqnarray}
\mu _1^2&=&-\frac{f_2}{\sqrt{2}}\frac{\upsilon _2 \upsilon _{\chi}}{\upsilon _1}-\lambda _1 \upsilon _1^2-\frac{1}{2}(\lambda _5+\lambda '_5)\upsilon _2^2-\frac{1}{2}\lambda _6\upsilon _{\chi }^2, \nonumber \\
\mu _2^2&=&-\frac{f_2}{\sqrt{2}}\frac{\upsilon _1 \upsilon _{\chi}}{\upsilon _2}-\lambda _2 \upsilon _2^2-\frac{1}{2}(\lambda _5+\lambda '_5)\upsilon _1^2-\frac{1}{2}\lambda _7\upsilon _{\chi }^2, \nonumber \\
\mu _3^2&=&-\frac{f_2}{\sqrt{2}}\frac{\upsilon _1\upsilon _2}{\upsilon _{\chi }}-\lambda _3 \upsilon _{\chi }^2-\frac{1}{2}\lambda _6\upsilon _1^2-\frac{1}{2}\lambda _{7 }\upsilon _2^2.
\end{eqnarray}
With the above relations replaced in the potential, we can obtain the squared mass matrices. Since the singlet $\sigma _0$ decouples from the other sectors at the quadratic order, we do not take into account this particle for the mass mixing. First, for the charged sector in the basis ${\phi _1^+,\phi _2^+}$, we find

\begin{eqnarray}
M_C^2&=&\mathcal{M}_{C}^2 \begin{pmatrix}
-T_{\beta} & 1
 \\
1 & -\left(T_{\beta }\right)^{-1}  \\
\end{pmatrix},
\label{charge-scalar-mass} 
\end{eqnarray}
where:

\begin{eqnarray}
\mathcal{M} _{C}^2&=&\frac{f_2 \upsilon _{\chi }}{\sqrt{2} }+\frac{\lambda '_5T_{\beta}}{2(1+T_{\beta}^2)}\upsilon ^2.
\label{charge-constant}
\end{eqnarray}
In the above expressions, we defined the following parameters:

\begin{eqnarray}
\upsilon ^2&=&\upsilon _{1}^2+ \upsilon _{2}^2, \nonumber \\ 
\tan \beta&=&T_{\beta}=\frac{\upsilon _2}{\upsilon _1}.
\label{parameters}
\end{eqnarray}
For the neutral real sector in the basis ${\xi _1,\xi _2, \xi _{\chi}}$, written as

\begin{eqnarray}
M_R^2&=&2\begin{pmatrix}
\left(\mathcal{M} _{R}^2\right)_{11} & \left(\mathcal{M} _{R}^2\right)_{12} & \left(\mathcal{M} _{R}^2\right)_{13}
 \\
* & \left(\mathcal{M} _{R}^2\right)_{22} &  \left(\mathcal{M} _{R}^2\right)_{23} \\
* & * &  \left(\mathcal{M} _{R}^2\right)_{33}  \\
\end{pmatrix},
\label{real-scalar-mass}
\end{eqnarray}
we find:

\begin{eqnarray}
\left(\mathcal{M} _{R}^2\right)_{ij}&=&\left(1-2\delta _{ij}\right)\frac{f_2\upsilon _1\upsilon _2\upsilon _{\chi }}{2\sqrt{2}\upsilon _i \upsilon _j}+\hat{\lambda }_{ij}\upsilon _i\upsilon _j,
\label{real-constant}
\end{eqnarray}
where the indices $i,j$ label the scalar indices $1,2 $ and $\chi =3 $. The symbol $\delta _{ij}$ is the Kronecker's delta. Also, we use for the couplings the following definitions:

\begin{eqnarray}
&&\hat{\lambda }_{11}=\lambda _1, \ \ \hat{\lambda }_{22}=\lambda _2, \ \ \hat{\lambda }_{33}=\lambda _3 \nonumber \\
&&\hat{\lambda }_{12}=\frac{1}{2}\left(\lambda _5+\lambda '_5\right), \ \ \hat{\lambda }_{13}=\frac{1}{2}\lambda _6, \ \ \hat{\lambda }_{23}=\frac{1}{2}\lambda _7. 
\end{eqnarray}
Finally, for the neutral imaginary sector, in the basis ${\zeta _1,\zeta _2, \zeta _{\chi}}$ and the form

\begin{eqnarray}
M_I^2&=&2\begin{pmatrix}
\left(\mathcal{M} _{I}^2\right)_{11} & \left(\mathcal{M} _{I}^2\right)_{12} & \left(\mathcal{M} _{I}^2\right)_{13}
 \\
* & \left(\mathcal{M} _{I}^2\right)_{22} &  \left(\mathcal{M} _{I}^2\right)_{23} \\
* & * &  \left(\mathcal{M} _{I}^2\right)_{33}  \\
\end{pmatrix},
\label{imag-scalar-mass}
\end{eqnarray}
we find:

\begin{eqnarray}
\left(\mathcal{M} _{I}^2\right)_{ij}&=&\left(1-2\delta _{ij}\right)\frac{f_2\upsilon _1\upsilon _2\upsilon _{\chi }}{2\sqrt{2}\upsilon _i \upsilon _j}.
\label{imag-constant}
\end{eqnarray}
It is interesting to observe that by comparing Eq. (\ref{real-constant}) and (\ref{imag-constant}), the real and imaginary components relate to each other according to:

\begin{eqnarray}
\left(\mathcal{M} _{R}^2\right)_{ij}-\left(\mathcal{M} _{I}^2\right)_{ij}&=&\hat{\lambda }_{ij}\upsilon _i\upsilon _j.
\label{real-imag}
\end{eqnarray}
In particular, for $i$ and $j=1,2$, the difference between both neutral sectors appears only at the electroweak order. Thus, some particles of the neutral heavy sector become degenerated if we take only the dominant term $\left|f_2\upsilon _{\chi}\right| \gg \upsilon ^2 $, which is also the scale of the charged sector. After diagonalization, we obtain the following mass eigenvectors:

\begin{eqnarray}
\begin{pmatrix}
G^{\pm}\\
H^{\pm}
\end{pmatrix}&=&R_{\beta}\begin{pmatrix}
\phi _1^{\pm}\\
\phi _2^{\pm}
\end{pmatrix}, \ \ \  
\begin{pmatrix}
G_{0}\\
A_{0}
\end{pmatrix}=R_{\beta}\begin{pmatrix}
\zeta _1\\
\zeta _2
\end{pmatrix}, \nonumber \\ 
\begin{pmatrix}
h_{0}\\
H_{0}
\end{pmatrix}&=&R_{\alpha}\begin{pmatrix}
\xi _1\\
\xi _2
\end{pmatrix}, \ \ \
\begin{pmatrix}
H_{\chi }\\
G_{\chi }
\end{pmatrix}\sim I\begin{pmatrix}
\xi _{\chi }\\
\zeta _{\chi }
\end{pmatrix}, 
\label{scalar-eigenvectors}
\end{eqnarray}
where $I$ is the identity, and the rotation matrices are defined according to

\begin{eqnarray}
R_{\beta, \alpha}&=&\begin{pmatrix}
C_{\beta, \alpha} & S_{\beta, \alpha} \\
-S_{\beta, \alpha} & C_{\beta, \alpha}
\end{pmatrix},
\end{eqnarray}
The rotation angles are the angle $\beta $ defined in (\ref{parameters}) and $\alpha $ obtained from the elements of the real matrix (\ref{real-scalar-mass}) as
\begin{eqnarray}
\tan{2\alpha }&\approx & \frac{2 \left(\mathcal{M} _{R}^2\right)_{12}}{\left(\mathcal{M} _{R}^2\right)_{11}-\left(\mathcal{M} _{R}^2\right)_{22}} \nonumber \\
&\approx & \tan{2\beta}\left[1+2\sqrt{2}S_{\beta}C_{\beta}\left(\frac{\hat{\lambda }_{22}T_{\beta}^2 - \hat{\lambda }_{11}}{T_{\beta}^2-1}\right)\left(\frac{\upsilon ^2}{f_2\upsilon _{\chi}}\right) \right]^{-1}
\label{alpha-angle}
\end{eqnarray}
where we have taken the dominant contribution assuming that $\upsilon ^2 \ll \left|f_2 \upsilon _{\chi}\right|$. We emphasis that although the second term into the inverse of (\ref{alpha-angle}) is proportional to the small ratio $\upsilon ^2 / f_2 \upsilon _{\chi}$, it is not negligibly small if $T_{\beta}^2 \approx 1$  and $\hat{\lambda }_{11} \neq \hat{\lambda }_{22}$. As for the eigenvalues, we obtain for the physical particles the following squared masses at dominant order:

\begin{eqnarray}
M_{H^{\pm}}^2&\approx&M_{H_{0}}^2\approx M_{A_{0}}^2\approx-\frac{f_2\upsilon _{\chi}}{\sqrt{2}}\left(\frac{1+T_{\beta}^2}{T_{\beta}}\right) \nonumber \\
M_{H_{\chi }}^2 &\approx &2\hat{\lambda }_{33} \upsilon _{\chi} ^2, \nonumber \\
M_{h_0}^2 &\approx& \frac{2\upsilon ^2}{\left(1+T_{\beta }^2\right)^2}\left(\hat{\lambda }_{11}+2\hat{\lambda }_{12}T_{\beta }^2+\hat{\lambda }_{22}T_{\beta }^4\right).
\label{scalar-mass}
\end{eqnarray}

From the above results, it is interesting to observe the following properties:

\begin{enumerate}

\item In Eq. (\ref{alpha-angle}), we can use the approximation $\tan{2\alpha} \approx \tan{2\beta}$ as dominant contribution only if the scalar doublets $\phi _{1,2}$ exhibit either 
  \begin{itemize}
  \item[-] the same self-interacting couplings, i.e. $\hat{\lambda }_{11}=\hat{\lambda }_{22}$, or 
  \item[-] different VEVs, i.e. $T_{\beta} \neq 1$.
  \end{itemize}

\item As we predicted below Eq. (\ref{real-imag}), the fields with masses given by the first line of  (\ref{scalar-mass}) are degenerated at dominant order.

\item In order to obtain masses with real values for the above degenerated particles, we require that either $f_2 < 0$ or $T_{\beta } < 0$.

\end{enumerate} 

On the other hand, we obtain all the couplings of the DM scalar $\sigma _0$ with the above mass eigenstates. The sector of the potential associated to $\sigma _0$ is:

\begin{eqnarray}
V_{\sigma}&=& \mu_{4}^2\left|\sigma _0\right|^2+\lambda _4 \left|\sigma _0\right|^4+\lambda '_{6} \left|\sigma _0\right|^2\left|\phi _1\right|^2+\lambda '_{7} \left|\sigma _0\right|^2\left|\phi _2\right|^2+\lambda _{8} \left|\sigma _0\right|^2\left|\chi _0\right|^2.
\end{eqnarray}  
After rotation to mass eigenvectors according to (\ref{scalar-eigenvectors}), we obtain all the interactions of $\sigma _0$ with the scalar matter:

\begin{eqnarray}
V_{\sigma}&=&M_{\sigma}^2\left|\sigma _0\right|^2-\upsilon \left(\lambda '_6S_{\alpha }C_{\beta }-\lambda '_7C_{\alpha }S_{\beta }\right)H_0\left|\sigma _0\right|^2 \nonumber \\
&+&\upsilon \left(\lambda '_6C_{\alpha }C_{\beta }+\lambda '_7S_{\alpha }S_{\beta }\right)h_0\left|\sigma _0\right|^2+\upsilon _{\chi }\left(\lambda _8\right)H_{\chi }\left|\sigma _0\right|^2 \nonumber \\
&+& \lambda _4\left|\sigma _0\right|^4+ \left(\lambda '_6S_{\beta } ^2+\lambda '_7C_{\beta }^2\right)\left|H^{+}\right|^2\left|\sigma _0\right|^2 \nonumber \\
&+&\frac{1}{2}\left(\lambda '_6S_{\alpha } ^2+\lambda '_7C_{\alpha }^2\right)\left(H_{0}\right)^2\left|\sigma _0\right|^2+\frac{1}{2}\left(\lambda '_6C_{\alpha } ^2+\lambda '_7S_{\alpha }^2\right)\left(h_{0}\right)^2\left|\sigma _0\right|^2 \nonumber \\
&-&S_{\alpha }C_{\alpha }\left(\lambda '_6-\lambda '_7\right)h_0H_0\left|\sigma _0\right|^2+\frac{1}{2}\left(\lambda '_6S_{\beta } ^2+\lambda '_7C_{\beta }^2\right)\left(A_{0}\right)^2\left|\sigma _0\right|^2+\frac{1}{2}\lambda _8\left(H_{\chi }\right)^2\left|\sigma _0\right|^2, 
\label{sigma-couplings}
\end{eqnarray}
where the mass of $\sigma _0$ is:

\begin{equation}
M_{\sigma }^2=\mu _{4}^2+\frac{1}{2}\left(\lambda '_6C_{\beta }^2+\lambda '_7S_{\beta }^2\right)\upsilon ^2+\frac{1}{2}\lambda _8\upsilon _{\chi } ^2.
\end{equation}

\subsection{Kinetic sector of the Higgs Lagrangian}

The kinetic terms of the Higgs Lagrangian reads:

\begin{eqnarray}
\mathcal{L}_{kin} &=&\sum_i (D_{\mu}S)^{\dagger}(D^{\mu}S).
\label{higgs-kinetic}
\end{eqnarray}
The covariant derivative is defined as

\begin{eqnarray}
D^{\mu}=\partial ^{\mu }-igW^{\mu}_{\alpha} T^{\alpha }_S-ig'\frac{Y_S}{2}B^{\mu}-ig_XX_SZ'^{\mu},
\label{covariant}
\end{eqnarray}
where $2T^{\alpha }_S$ corresponds to the Pauli matrices for the doublets $S=\phi _{1,2}$ and $T^{\alpha }_S=0$ for the singlets  $S=\chi _0, \sigma _0$, while $Y_S$ and $X_S$ corresponds to the hypercharge and $U(1)_X$ charge according to values of Tabs. \ref{tab:SM-espectro} and \ref{tab:exotic-espectro}. After the symmetry breaking, we obtain the charged eigenstates $W_{\mu }^{\pm}=(W_{\mu }^{1}\mp W_{\mu }^{2})/\sqrt{2}$ with mass $M_{W}=g\upsilon /2$, while for the neutral sector, we obtain the following squared mass matrix in the neutral gauge basis  $({W_{\mu }^3,B_{\mu},Z'_{\mu}})$:

\begin{eqnarray}
M_0^2&=&\frac{1}{4}\begin{pmatrix}
g^2\upsilon ^2 &-gg'\upsilon ^2   &-\frac{2}{3}gg_X\upsilon ^2(1+C_{\beta} ^2)   \\ 
&&\\
* &  g'^2\upsilon ^2 &  \frac{2}{3}g'g_X\upsilon ^2(1+C_{\beta}^2)   \\
 &&\\
* &*  &  \frac{4}{9}g_X^2\upsilon _{\chi}^2\left[1+(1+3C_{\beta}^2)\epsilon^2\right] \\
\end{pmatrix},
\end{eqnarray}
where $\epsilon =\upsilon /\upsilon _{\chi }$. Taking into account that $\epsilon ^2 \ll 1$, we can diagonalize the above matrix with only two rotation angles, obtaining the following mass eigenstates:

\begin{eqnarray}
\begin{pmatrix}
A_{\mu} \\
Z_{1\mu } \\
Z_{2\mu } 
\end{pmatrix} &\approx &R_0
\begin{pmatrix}
W_{\mu}^3 \\
B_{\mu } \\
Z'_{\mu } 
\end{pmatrix},
\label{gauge-eigenvec}
\end{eqnarray}  
with:

\begin{eqnarray}
R_0&=&\begin{pmatrix}
S_W & C_W  & 0 \\ 
&&\\
C_WC_{\theta} &  -S_WC_{\theta}  &  S_{\theta}   \\
 &&\\
-C_WS_{\theta}  & S_WS_{\theta}   & C_{\theta}  \\
\end{pmatrix},
\end{eqnarray}
where $\tan{\theta _W}=S_W/C_W=g'/g$ defines the Weinberg angle, and $S_{\theta }=\sin{\theta }$ is a small $Z$-$Z'$ mixing angle between the SM neutral gauge boson $Z$ and the $U(1)_X$ gauge boson $Z'$ such that in the limit $S_{\theta}=0$, we obtain $Z_1=Z$ and $Z_2=Z'$. We find for the mixing angle that

\begin{eqnarray}
S_{\theta} \approx  (1+C_{\beta}^2)\frac{2g_XC_W}{3g}\left(\frac{M_Z}{M_{Z'}}\right)^2,
\label{mixing-angle}
\end{eqnarray}
where the neutral masses are:

\begin{eqnarray}
M_Z&\approx& \frac{g\upsilon }{2C_W} ,\ \ \ \  \  \ M_{Z'}\approx  \frac{g_X\upsilon _{\chi }}{3}.
\end{eqnarray}

On the other hand, the Lagrangian in (\ref{higgs-kinetic}) also contains the interactions among scalar and vector bosons. In particular, for the DM candidate $\sigma _0$ and considering the mixing $Z$-$Z'$ angle, we find the following couplings:

\begin{eqnarray}
\mathcal{L}_{\sigma } &\approx& \frac{1}{3}g_XS_{\theta}\left(q-p\right)_{\mu } Z^{\mu }_1\left|\sigma _0\right|^2+\frac{1}{3}g_X\left(q-p\right)_{\mu } Z^{\mu }_2\left|\sigma _0\right|^2 \nonumber \\
&+&\frac{1}{9}g_{X}^2S_{\theta}^2\left|Z_1\right|^2\left|\sigma _0\right|^2+\frac{2}{9}g_{X}^2S_{\theta}Z^{\mu }_1Z_{2\mu }\left|\sigma _0\right|^2+\frac{1}{9}g_{X}^2\left|Z_2\right|^2\left|\sigma _0\right|^2
\label{gauge-interaction}
\end{eqnarray}
where we approximate $C_{\theta}\approx 1$. In the trilineal terms, we introduce the vectors $q_{\mu}$ ($p_{\mu}$) corresponding to the $\sigma _0$ ($Z_{1,2}$) momentum.

\subsection{Neutral weak couplings}

The couplings between fermions and gauge bosons, are described by the Dirac Lagrangian, which reads:

\begin{eqnarray}
\mathcal{L}_{D}=i\sum_{f,i}\overline{f^i_L}\gamma ^\mu D_{\mu} f^i_L+\overline{f^i_R}\gamma ^\mu D_{\mu}f^i_R, 
\end{eqnarray}
where $f^i_{L,R}$ contains SM and non-SM fermions. In particular, for the neutral weak sector, and taking into account the mass states in (\ref{gauge-eigenvec}), we obtain \cite{U1}

\begin{eqnarray}
\mathcal{L}_{WN}&=&Z_{1\mu }\sum _{f,i}\overline{f^i}\gamma ^{\mu}\left[\frac{g}{2C_W}(v_i^{SM}-\gamma _5a_i^{SM})C_{\theta}-\frac{g_X}{2}(v_i^{NSM}-\gamma _5a_i^{NSM})S_{\theta}\right]f^i \nonumber \\
&-&Z_{2\mu }\sum _{f,i}\overline{f^i}\gamma ^{\mu}\left[\frac{g}{2C_W}(v_i^{SM}-\gamma _5a_i^{SM})S_{\theta}+\frac{g_X}{2}(v_i^{NSM}-\gamma _5a_i^{NSM})C_{\theta}\right]f^i, 
\label{weak current}
\end{eqnarray} 
where the vector and axial couplings for the weak neutral currents are defined according to Tab. \ref{tab:vector-axial-couplings}. We observe that a small non-SM coupling ($g_X$) with the lighter weak boson $Z_1$ arises through the mixing angle $S_{\theta}$. The above Lagrangian can be written in a shorter form by defining rotations to modified vector and axial couplings as 

\begin{eqnarray}
\begin{pmatrix}
\overline{v}_{i}^{(1)} \\
\\
\overline{v}_{i}^{(2)}
\end{pmatrix}&=&
\mathcal{R}\begin{pmatrix}
v_i^{SM} \\
\\
v_i^{NSM}
\end{pmatrix}, \ \ \ \ \begin{pmatrix}
\overline{a}_{i}^{(1)} \\
\\
\overline{a}_{i}^{(2)}
\end{pmatrix}=
\mathcal{R}\begin{pmatrix}
a_i^{SM} \\
\\
a_i^{NSM}
\end{pmatrix}
\label{modified-coup}
\end{eqnarray}
where:

\begin{eqnarray}
\mathcal{R}=\frac{g_X}{g}\begin{pmatrix}
\frac{g}{g_X}C_{\theta}& -C_WS_{\theta} \\
& \\
\frac{g}{g_X}S_{\theta} & C_WC_{\theta}
\end{pmatrix},
\end{eqnarray}
obtaining

\begin{eqnarray}
\mathcal{L}_{WN}&=&\frac{g}{2C_W}\sum _{f,i}\left[\overline{f^i}\gamma ^{\mu}\left(\overline{v}_{i}^{(1)}-\gamma _5\overline{a}_{i}^{(1)} \right)f^iZ_{1\mu }+\overline{f^i}\gamma ^{\mu}\left(\overline{v}_{i}^{(2)}-\gamma _5\overline{a}_{i}^{(2)} \right)f^iZ_{2\mu } \right]
\label{weak-neutral-lag}
\end{eqnarray}

\subsection{Yukawa Lagrangian}

We obtain the Yukawa Lagrangian compatible with the $G_{sm}\times U(1)_X$ symmetry. For the quark sector we find:

\begin{eqnarray}
-\mathcal{L}_Q &=& \overline{q_L^{a}}(\widetilde{\phi }_1 h^{U}_{1})_{aj}U_R^{j}+ \overline{q_L^{3}}\left(\widetilde{\phi} _2h^{U}_2 \right)_{3j}U_R^{j}+\overline{q_L^{a}}\left(\phi _2 h^{D}_{2} \right)_{aj}D_R^{j}+\overline{q_L^{3}}\left(\phi _1 h^{D}_1\right)_{3j}D_R^{j} \notag \\
&+&\overline{q_L^{a}}\left(\phi  _2 h^{J}_{2} \right)_{am} J^{m}_R+ \overline{q_L^{3}} (\phi _1 h^{J}_{1})_{3m} J^{m}_R+\overline{q_L^{a}} (\widetilde{\phi } _1 h^{T}_{1})_aT_R+\overline{q_L^{3}}\left(\widetilde{\phi} _2 h^{T}_{2} \right)_3T_R  \notag \\
&+&\overline{T_{L}}\left( \sigma_0 h_{\sigma }^{U}+\chi _0h_{\chi }^{U}\right)_{j}{U}_{R}^{j}+\overline{T_{L}}\left( \sigma_0h_{\sigma}^{T}+\chi _0h_{\chi }^{T}\right){T}_{R}
\nonumber \\
&+&\overline{J_{L}^n}\left( \sigma_0^*h_{\sigma }^{D}+\chi _0^*h_{\chi }^{D}\right)_{nj}{D}_{R}^{j}+\overline{J_{L}^n}\left( \sigma_0^*h_{\sigma }^{J}+\chi _0^*h_{\chi }^{J}\right)_{nm}{J}_{R}^{m}+h.c.,
 \label{quark-yukawa-1}
\end{eqnarray}
where $\widetilde{\phi}_{1,2}=i\sigma_2 \phi_{1,2}^*$ are conjugate scalar doublets, and $a=1,2$. 
For the leptonic sector we obtain:

\begin{eqnarray}
-\mathcal{L}_{\ell}&=&\overline{\ell ^{i}_{L}}\left( \widetilde{\phi}_1h_{1 }^{\nu } \right)_{ij}{\nu }_{R}^{j}+ \overline{\ell ^{i}_{L}}\left( \widetilde{\phi}_2h_{2 }^{N}\right)_{ij}{N}_{R}^{j}\nonumber \\
&+&\overline{(\nu ^{i}_{R})^c}\left( \sigma _0^*h_{\sigma }^{N}+\chi _0^* h_{\chi }^{N}\right)_{ij}{N}_{R}^j +\frac{1}{2}M_{N}\overline{(N^{i}_{R})^c}{N}_{R}^{j}\nonumber \\
&+&\overline{\ell ^{i}_{L}}\left( \phi _1h_{1}^{e } \right)_{ij}{e}_{R}^{j}+h.c.
\label{yukawa-leptons-1}
\end{eqnarray}
In particular, we can see in the quark Lagrangian from Eq. (\ref{quark-yukawa-1}) that due to the non-universality of the $U(1)_X$ symmetry, not all couplings between quarks and scalars are allowed by the gauge symmetry, which lead us to specific zero-texture Yukawa matrices as studied in ref. \cite{U1}. For the purpose of this paper, we can assume diagonal Yukawa matrices. In addition, if the global symmetry in (\ref{global-symm}) is required only for $\sigma _0$, its couplings with fermions are forbidden, thus we assume that $h_{\sigma }^f=0$ in (\ref{quark-yukawa-1}) and (\ref{yukawa-leptons-1}). However, the $\sigma _0$ particle can still scatter off to fermions indirectly through processes of the form $\sigma_0\sigma _0^*\rightarrow \phi _i \rightarrow f\overline{f}$, where the decays of the scalar particles into ordinary matter are described by the couplings in the first line of (\ref{quark-yukawa-1}) and the last of (\ref{yukawa-leptons-1}). By considering the dominant contributions, we can decouple the heavy and the light sectors of the fermions to obtain the mass eigenvalues after the symmetry breaking. In particular, with the above assumptions and under the assignation from (\ref{quark-assignation}), the Yukawa couplings of scalars to ordinary matter is the same as in a two Higgs doublet model type III. Since these models contains contributions to flavor changing neutral currents, which are severely suppresed by experimental data at electroweak scales, we will assume that the Yukawa couplings relate to the mass of quarks in the same way as in a type II model, avoiding the introduction of unnecesary additional free parameters.

\section{Z Pole constraints}

As a result of the mixing in the couplings with the weak boson $Z_1$ in (\ref{weak-neutral-lag}), small deviations of the SM observable arise. In particular, we can use the Z pole observables shown in Tab. \ref{tab:observables} with their experimental values from the CERN collider (LEP), the SLAC Liner Collider (SLC) and data from atomic parity violation \cite{data}, and the theoretical predictions from SM and the deviations predicted by the $U(1)_X$ model, as shown in the appendix \ref{app:z-pole}. We use $M_Z=91.1876$ GeV and $S_W^2=0.23108$. All the technical details including the electroweak corrections considered are summarized in the appendix. We perform a $\chi ^2$ fit at $95\%$ C.L., where the free quantities $T_{\beta}, M_{Z'}$ and the ratio $r_g=g_X/g$ can be constrained at the Z peak. We assume a covariance matrix with elements $V_{ij}=\rho _{ij}\sigma
_{i}\sigma _{j}$ among the Z pole observables, where $\rho $ is the correlation
matrix and $\sigma $ the quadratic root between the experimental and SM errors.
The $\chi ^{2}$ statistic with three degrees of freedom (d.o.f) is defined
as \cite{data}

\begin{equation}
\chi ^{2}(T_{\beta },M_{Z'},r_g )=\left[ \mathbf{y}-\mathbf{F}%
(T_{\beta },M_{Z'},r_g )\right] ^{T}V^{-1}\left[ \mathbf{y}-\mathbf{F}%
(T_{\beta },M_{Z'},r_g )\right] ,  \label{chi}
\end{equation}
where $\mathbf{y=\{}y_{i}\mathbf{\}}$ represents the 22 experimental observables from table \ref{tab:observables}, and $\mathbf{F}$ the corresponding prediction from the $U(1)_X$ model. Table \ref{tab:correlation} displays the symmetrical correlation matrices taken from ref. \cite{LEP}. The function in (\ref{chi}) imposes constraints to the free variables ($T_{\beta
},M_{Z'},r_g $)  by requiring that $\chi ^{2}\leq \chi _{\min}^{2}+\chi _{CL}^{2}$.
For three d.o.f the $95\%$ C.L. corresponds to $\chi _{CL}^{2}=7.815$. With the family assignation in (\ref{quark-assignation}) and the Z pole parameters, we find for the minimum that $\chi _{\min }^{2}=156.298$.

\begin{figure}[tbh]
\centering
\includegraphics[width=11cm,height=8cm,angle=0]{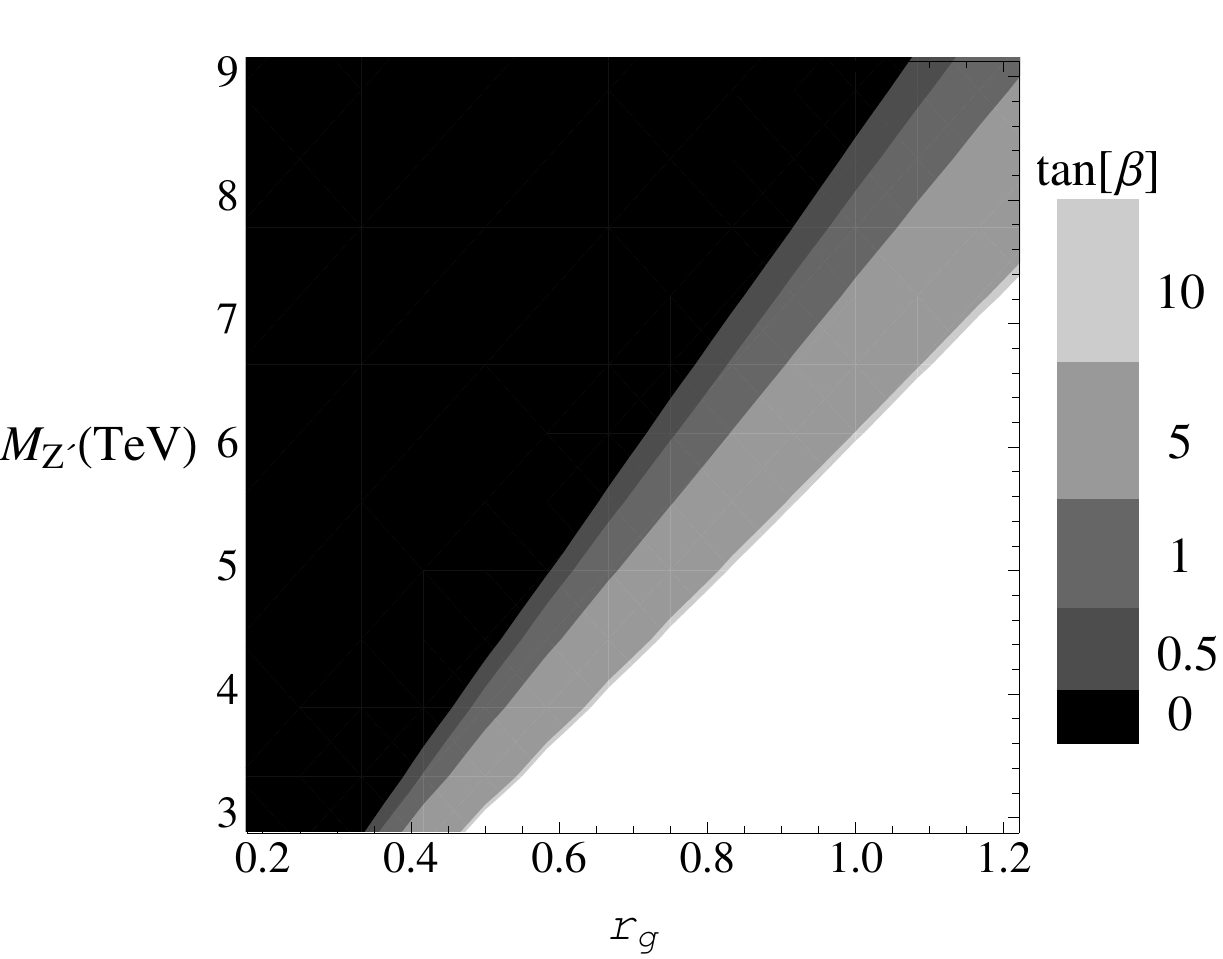}
\caption{Constraints for $M_{Z'}$ in the $(M_{Z'}, r_{g})$ plane for different values of $T_{\beta }$, with $r_{g}=g_X/g$ the ratio between the $U(1)_X$ and $SU(2)_L$ gauge couplings. The shaded areas show the allowed points from Z pole constraints.}
\label{fig1}
\end{figure}

We obtain two-dimensional projections of the $\chi ^2$ function. Fig. \ref{fig1} shows the limits for the $Z'$ mass in the ($M_{Z'}, r_g$) plane. The shaded areas correspond to allow regions for different values of $T_{\beta}$, where we identify bounds for $M_{Z'}$ and the ratio $r_g$. Thus, for example, around the experimental lowest limit of $M_{Z'} = 3$ TeV, we see that the model is allowed if $g_X/g\leq 0.35, 0.4$ and $0.5$ when $T_{\beta}=0, 1$ and $10$, respectively. These limits on the coupling constant ratio increase for larger $Z'$ masses. In particular, for a model with $g_X=g$,  we find that $M_{Z'}$ must be above $6, 7.4$ and $8.5$ TeV if $T_{\beta}=10, 1$ and $0$, respectively.  

\begin{figure}[tbh]
\centering
\includegraphics[width=11cm,height=8cm,angle=0]{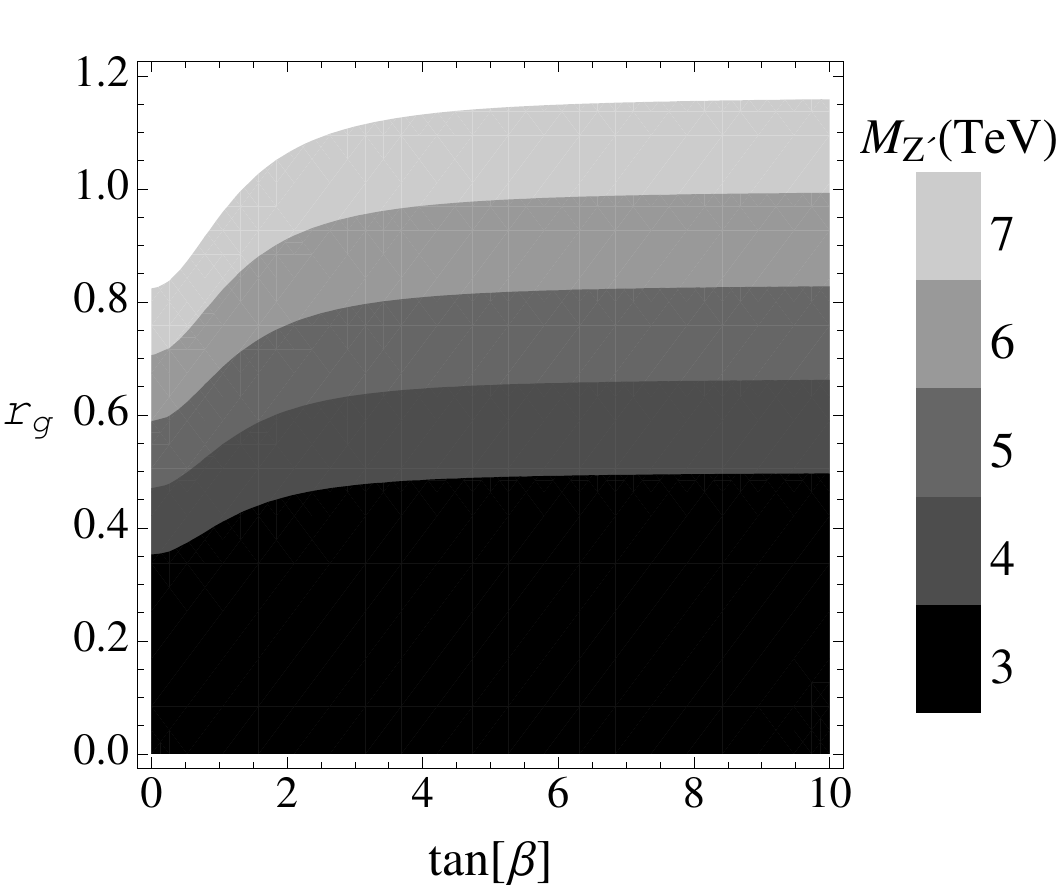}
\caption{Constraints for $r_{g}$ in the $(r_{g},T_{\beta})$ plane for different values of $M_{Z' }$, with $r_{g}=g_X/g$ the ratio between the $U(1)_X$ and $SU(2)_L$ gauge couplings. The shaded areas show the allowed points from Z pole constraints}
\label{fig2}
\end{figure}

On the other hand, Fig. \ref{fig2} shows the constraints on $r_g$ as function of $T_{\beta}$. We observe that for $M_{Z'}=3$ TeV, the model is completely excluded for $g_X/g > 0.5$. We also see that for a $3$ TeV $Z'$ boson, values near $T_{\beta}=0$ are not allowed in the range $0.35 \leq g_X/g \leq 0.5$. In the scenario with $g_X=g$, the range $\left|T_{\beta}\right| \leq 1.3$ is excluded for a $7$ TeV $Z'$ boson, while for a $6$ TeV gauge boson, the model exhibits allowed points only at large $T_{\beta}$ values. We found that the above constraints do not change significantly for values above $T_{\beta}=10$.

\section{Constraints from the relic density}

The Boltzmann equation for the ratio of the number density for DM particles is \cite{boltz}:

\begin{eqnarray}
\frac{dn(t)}{dt}=-3Hn(t)-\langle v_{rel}\sigma _{ann} \rangle \left(n(t)^2-n_{eq}^2\right),
\end{eqnarray}  
where three sources for variation of the density are identified: first, a negative contribution due to the expansion of the Universe described by the Hubble constant ($H$). Second, there is another negative contribution coming from the annihilation of DM into ordinary matter. This contribution is described by a thermally-averaged pair-annihilation cross sections times the relative velocity between the DM particles $\langle v_{rel}\sigma _{ann} \rangle$, and the instantaneous squared density $n(t)^2$ of DM (the higher the density, the greater the annihilation probability). Finally, there is a positive contribution which describes DM creation through matter collisions, which exhibits the same thermal cross section. This contribution depends on the number of particles in thermal equilibrium ($n_{eq}$). The Boltzmann equation is conveniently written in terms of the abundance $Y=n/s$ and the temperature $T$, with $s$ the entropy of the Universe, obtaining:

\begin{eqnarray}
\frac{dY(T)}{dT}=\sqrt{\frac{\pi g_{*}(T)}{45}}M_p\langle v_{rel}\sigma _{ann} \rangle \left(Y(T)^2-Y_{eq}^2\right),
\end{eqnarray}
where $M_p$ is the Planck mass and $g_{*}(T)$ the number of degrees of freedom in thermal equilibrium with the photons and with masses smaller than $T$. By solving the above equation, it is possible to calculate the relic density, which is defined as:

 \begin{eqnarray}
 \Omega _{DM}h^2=\frac{n_0M_{DM}}{\rho _{c}}h^2=\frac{s_0Y_0M_{DM}}{\rho _c}h^2,
 \end{eqnarray}
where $\rho _c$ is the critical density and $M_{DM}$ the mass of the DM candidate, while $s_0$ and $Y_0$ are the entropy and abundance evaluated at the today temperature $T_0=2.726$ K measured from the microwave background. In our case, we will consider that the complex $\sigma _0$ particle is the single DM component of the Universe, thus $M_{DM}=M_{\sigma}$ and the thermal cross section is obtained from all the dispersions of $\sigma _0$ according to the couplings obtained in section 2. To calculate the freeze-out temperature, the thermal cross section and the relic density, we use the numerical code MicrOMEGAs \cite{micromegas}, where we assume that the abundance accomplish the condition $Y_{eq}\approx Y(T)$ for high temperatures until freeze-out, and $Y_{eq}=0$ for temperatures below the freeze-out \cite{freeze-out}. For the thermal cross section, we identify the following parameters:

\begin{enumerate}

\item For direct dispersions into neutral gauge bosons,  $\sigma _0\sigma ^{*}_0 \rightarrow Z_1 (2Z_1)$ described by the couplings in Eq. (\ref{gauge-interaction}), we include the restrictions derived from the Z pole constraints for the parameters $T_{\beta}$, $g_X$ and $M_{Z'}$. In particular, according to Fig. \ref{fig1}, the larger value of the gauge coupling ratio is around $r_g\sim 0.4$ for a $3$ TeV extra neutral gauge boson. Thus, we reduce the number of parameters by choosing the above values, which are compatible with the Z pole constraints. According to Eq. (\ref{mixing-angle}), these values lead us to a $Z$-$Z'$ mixing angle as large as $S_{\theta}=0.4 \times 10^{-3}$.     

\item For indirect dispersions through the scalar particles $S$,  $\sigma _0\sigma ^{*}_0 \rightarrow S (2S) $ described by (\ref{sigma-couplings}), we can reduce the number of parameters by considering a model with two identically self-interacting scalar doublets, where $\hat{\lambda }_{11} =\hat{\lambda }_{22}=\hat{\lambda }_{12} $. This assumption lead us that the angle $\alpha $ in Eq. (\ref{alpha-angle}) become

\begin{eqnarray}
\tan{2\alpha }=\tan{2\beta }\left(1-\frac{M_{h_0}^2}{M_{H_0}^2}\right).
\label{alpha-angle-2}
\end{eqnarray}
Thus, if we assume that $h_0$ is the experimentally observed boson at LHC (where $M_{h_0}=126$ GeV), both angles are related by only one free parameter, i.e the mass of the neutral Higgs boson $H_0$. Since  no positive signal in the search for extra neutral Higgs bosons at LHC has been reported, we choose a large value $M_{H_{0}}=500$ GeV. The other free parameters in (\ref{sigma-couplings}) are the couplings $\lambda '_6$ and $\lambda '_7$. 
We recall that the pseudo scalar $A_0$ and the charged scalar $H^{\pm}$ have masses at the same order as $H_0$ at dominant order, according to Eq. (\ref{scalar-mass}).   

\item Using a two Higgs doublet model type II for the Yukawa couplings, the decay of the scalar matter into ordinary fermions $S \rightarrow f_{SM}\overline{f_{SM}}$ involve the fermion masses, the electroweak VEV and the angle $\beta$. Thus, these couplings do not introduce additional free parameters.

\item The decay of scalars into gauge bosons $S \rightarrow 2Z (2W^{\pm}, ZW^{\pm})$ only involve the known $SU(2)_L$ coupling constant $g$. Thus, these interactions do not introduce more free parameters neither.

  
\end{enumerate}

Thus, our space of parameters is composed by the four variables $(M_{\sigma}, T_{\beta},\lambda '_6, \lambda '_7)$. Taking into account that the current limit for DM relic density is $\Omega h^2=0.1199\pm0.0027$ from the Planck experiment \cite{planck}, we search for limits on the mass of the scalar DM candidate $M_ \sigma$ for different combinations of the parameters. We obtain the following results.

\begin{figure}[tbh]
\centering
\subfigure[]{\includegraphics[width=8cm,height=5cm,angle=0]{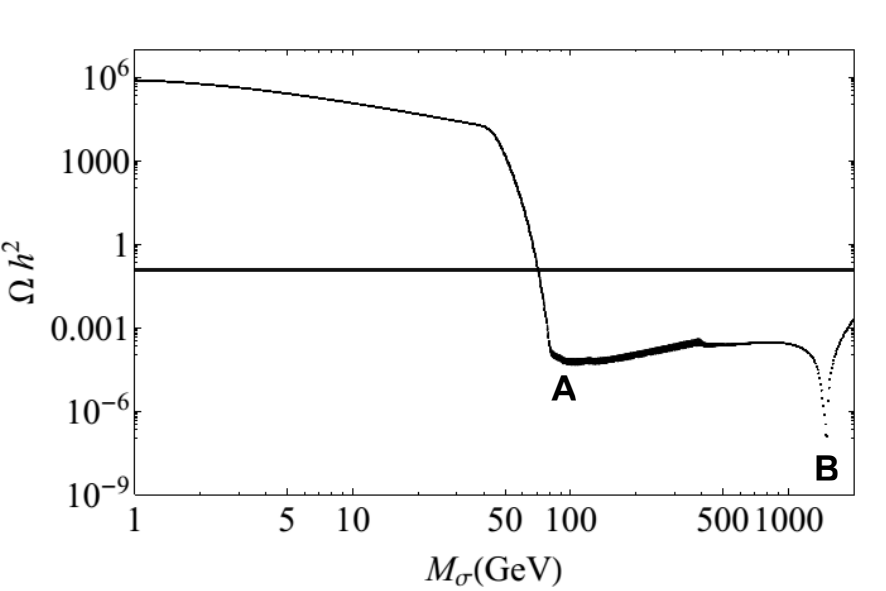}}\hspace{-0cm}
\subfigure[]{\includegraphics[width=8cm,height=5cm,angle=0]{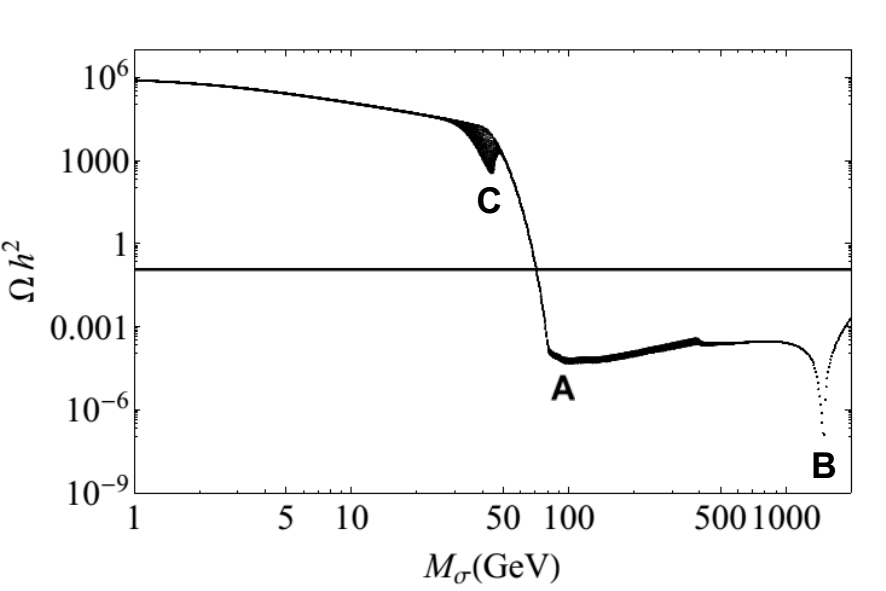}}
\caption{Relic density as function of $M_{\sigma}$ in the higgsphobic model with $\lambda '_{6,7}=0$. In (a) the gauge $Z$ boson decouple from DM, where $S_{\theta}=0$. In (b) a small resonance (\textbf{C}) arises due to the coupling of $Z_1$ to DM through the mixing angle, where $0 \leq T_{\beta} \leq 10$. \textbf{A} label the electroweak threshold due to couplings between the $Z'$ boson and the other weak gauge bosons. The peak at \textbf{B} is due to the resonance of the $Z'$ boson}
\label{fig3ab}
\end{figure}

\begin{figure}[tbh]
\centering
\includegraphics[width=11cm,height=8cm,angle=0]{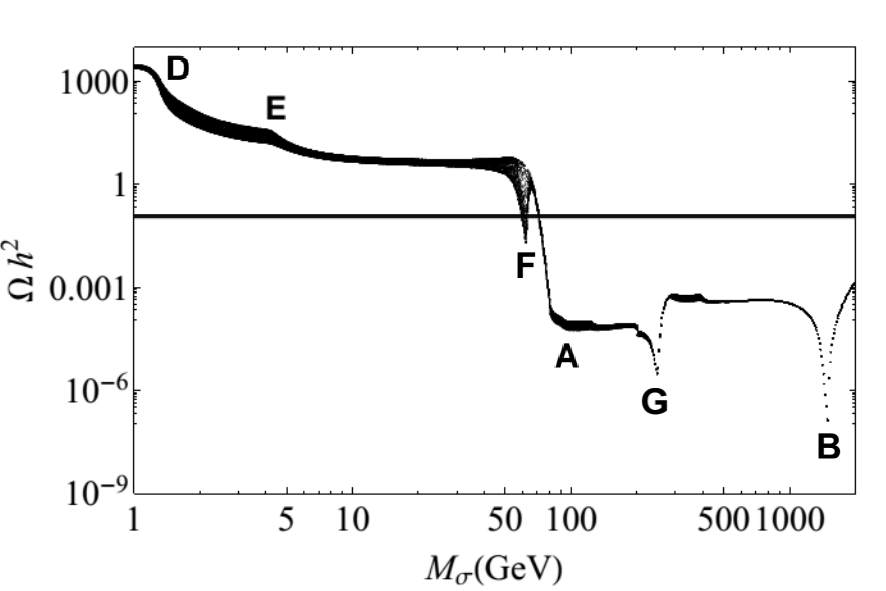}\vspace{-0cm}
\caption{Relic density as function of $M_{\sigma}$ with $\lambda '_{6,7}=1$ and $0\leq T_{\beta}\leq 1 $. The \textbf{D}, \textbf{E} and \textbf{A} signatures label kinematical thresholds. \textbf{F}, \textbf{G} and \textbf{B} are resonances due to intermediate particle production}
\label{fig4}
\end{figure}

\begin{figure}[tbh]
\centering
\includegraphics[width=11cm,height=8cm,angle=0]{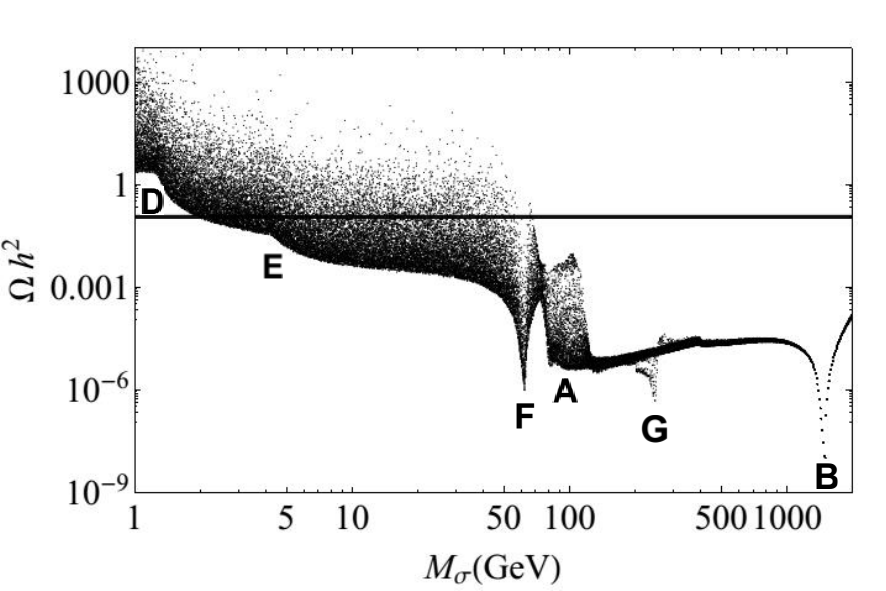}\vspace{-0cm}
\caption{Relic density as function of $M_{\sigma}$ with $0 \leq \lambda '_{6,7} \leq 3$ and $0\leq T_{\beta}\leq 10 $. The \textbf{D}, \textbf{E} and \textbf{A} signatures label kinematical thresholds. \textbf{F}, \textbf{G} and \textbf{B} are resonances due to intermediate particle production.}
\label{fig5}
\end{figure}

\begin{figure}[tbh]
\centering
\includegraphics[width=11cm,height=8cm,angle=0]{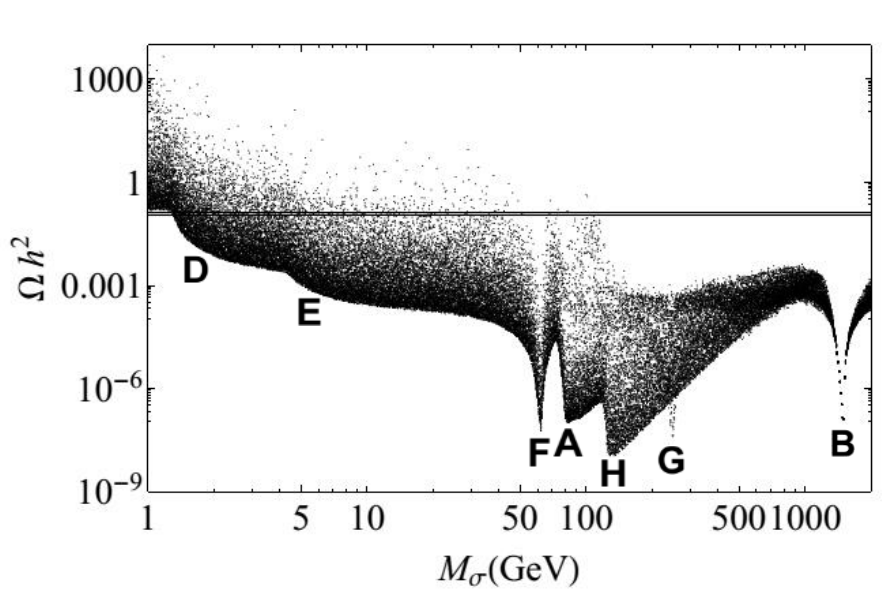}\vspace{-0cm}
\caption{Relic density as function of $M_{\sigma}$ with $0 \leq \lambda '_{6,7} \leq 4\pi$ and $0\leq T_{\beta}\leq 10 $. The \textbf{D}, \textbf{E}, \textbf{A} and \textbf{H} signatures label the kinematical thresholds. \textbf{F}, \textbf{G} and \textbf{B} are resonances due to intermediate particle production.}
\label{fig6}
\end{figure}

\begin{enumerate}
\item In order to explore the contribution of the gauge $Z_1$ and $Z_2$ bosons, we first consider the \emph{higgsphobic} case, by fixing $\lambda '_{6,7}=0$. Fig. \ref{fig3ab}-(a) shows the relic density as function of $M_{\sigma}$ without $Z$-$Z'$ mixing. According to Eq. (\ref{gauge-interaction}), the light $Z_1=Z$ gauge boson decouple from the DM when $S_{\theta}=0$, thus the dominant channel for DM annihilation is through the $Z_2=Z'$ gauge boson. We see that below $M_{\sigma }=70$ GeV, very large DM relic density is obtained which means that there are few DM annihilation through the freeze-out due to the small coupling to ordinary matter. However, when the mass of the $\sigma _0$ particle increases until the electroweak threshold at $M_{\sigma} \approx 91$ GeV, labeled in figure (a) as \textbf{A}, the probability of annihilation into ordinary matter grows. We see that at $M_{\sigma }=1500$ GeV (labeled as \textbf{B}), there arises a resonance due to the pole $(2M_{\sigma })^2-M_{Z'}^2=0$, which produces an extra excess of annihilation through the process $\sigma _0\sigma ^{*}_0 \rightarrow Z' \rightarrow SM,SM$. The horizontal line is the experimental limit for the relic density. Thus, for this case, the appropriate mass is $M_{\sigma }=70$ GeV to fit the data. In Fig. \ref{fig3ab}-(b) we take values for the mixing angle. In particular, we choose random values for the parameter $T_{\beta}$, which according to the Z pole limits gives values as large as $S_{\theta}=0.4\times 10^{-3}$ for $M_{Z'}=3$ TeV. The plot is the same as the curve in (a), but with a new small resonance at $M_{\sigma } \approx 45$ GeV (label \textbf{C}) corresponding to the pole $(2M_{\sigma })^2-M_{Z}^2=0$, which is consequence of the coupling of $Z$ bosons to DM. Although the relic density decreases at this resonance, the annihilation ratio is not large enough to reach the observable density. Thus, the Z pole data produce strong constraints to fit the relic density for higgsphobic models.

\item Let us take $\lambda '_{6,7}=1$, which open the scalar channels. We consider small values for $T_{\beta}$. In particular, we choose random values between $0$ and $1$, obtaining the plot from Fig. \ref{fig4}. First, we see for light DM that the relic density decreases in relation to the above case. Thus, the coupling with the scalar matter produces larger annihilation rates than the coupling with gauge bosons. In particular, for $M_{\sigma }=M_{Z}/2$, the scalar channels produce lower relic densities that the resonance of the $Z_1$ boson, thus this resonance does not appear in the plot. In more detail, we see that there are multiple kinematical thresholds through the curve with observable drops of the density. The first threshold happens at $M_{\sigma }\approx 1.3$ GeV (\textbf{D}), due to dispersions of $\sigma $ into charm quarks. Later, there is another drop at $M_{\sigma }\approx 4$ GeV (\textbf{E}) from production of bottom quarks. Finally, there is a third large drop due to dispersions into $W^{\pm}$ and $Z$ gauge bosons at $M_{\sigma }=80-91$ GeV (\textbf{A}). On the other hand, we also identify sharp resonances associated to production of intermediate scalar particles. The first at $M_{\sigma}=M_{h_{0}}/2=63$ GeV  (\textbf{F}) corresponding to the process $\sigma_ 0\sigma_ 0^{*}\rightarrow h_{0}\rightarrow  SM,SM$ through the SM-like Higgs boson with mass of $126$ GeV. We see that below the electroweak threshold, the density is above the experimental value, except for the resonance, that exhibits allowed points around $M_{\sigma}=63$ GeV. The other allow point is at $M_{\sigma}=70$ GeV which is the same as the one obtained in Figure \ref{fig3ab}. Above the electroweak threshold, the relic density drops below the observable value. In particular, we see that there are two others resonances at $M_{\sigma} =M_{H_{0}}/2=250$ GeV  (\textbf{G}) and $M_{\sigma}=M_{Z'}/2=1500$ GeV (\textbf{B}).

\item In Fig. \ref{fig5}, we show the relic density for random values in the ranges $0 \leq \lambda '_{6,7}\leq 3$ and $0 \leq T_{\beta}\leq 10$. We see that the general form of the plot from Fig. \ref{fig4} is preserved. However, solutions associated to large values of $T_{\beta}$ and $\lambda '_{6,7}$ appears through the horizontal line, as shown. The allow values for $M_{DM}$ is an important contraint for DM research in direct detection and production in collider experiments. On the other hand,, we see that the resonance due to SM-like Higgs boson production at \textbf{F} gives a relic density below the observable value, producing a deep loss of DM in the window $55 \text{ GeV} \leq M_{\sigma}\leq 75$ GeV. We also see that there are solutions thorugh the observable relic density in the range $2 \text{ GeV} \leq M_{\sigma} \leq 50$ GeV

\item Finally, we open our space of parameters by extending the $T_{\beta }$ between $0$ to $10$, and the $\lambda '_{6,7}$ between $0$ to the limit for unitarity of $4\pi$. Fig. \ref{fig6} shows the results for random values. We identify again the three old kinematical threshold $\textbf{D}$, $\textbf{E}$ and $\textbf{A}$ and the same resonances $\textbf{F}$, $\textbf{G}$ and $\textbf{B}$. In addition, another peak at $M_{\sigma}=126$ GeV appears (\textbf{H}), which is hardly visible in Fig. \ref{fig5}. This peak is not due to a new resonance but a fourth kinematical threshold from dispersions of $\sigma _0$ into Higgs bosons $h_0$ due to large coupling values $\lambda '_{6,7}$ that increases the dispersion rates into the Higgs. Roughly, we see that there exist values of the space of parameters that reproduce the experimental relic density for the range $1.3 \text{ GeV} \leq M_{\sigma} \leq 50$ GeV and  $65 \text{ GeV} \leq M_{\sigma} \leq 125$ GeV. By comparing Figs. \ref{fig5} and \ref{fig6}, we see that the larger the scalar couplings, the larger the allow ranges for the DM mass.          

\end{enumerate}

\section{Conclusions}

The puzzle of the DM in the Universe can be related with particles that exhibit additional non-gravitational interactions in the framework of particle physics theories. 
In particular, extensions with an extra non-universal abelian $U(1)'$ symmetry are very well-motivated models that can introduce different DM candidates with new weak interactions through the addition of a neutral gauge boson $Z'$. In this work, we studied some consequences of this neutral current in an specific $U(1)'$ model with two scalar doublets and two scalar singlets that exhibit nontrivial $U(1)'$ charges, where one of the singlet is taken as a DM candidate.

We first explore some consequences of the $Z$-$Z'$ mixing using data from LEP, SLAC and atomic parity violation. We found for the lower experimental limit $M_{Z'}=3$ TeV, that the $U(1)'$ coupling constant is constrained to values as large as $g_X=0.5g$ for $T_{\beta}\geq 10$ and at $95\%$ C.L. This limit increases for larger $M_{Z'}$ values. Second, the $Z$-$Z'$ mixing induces dispersions of the DM candidate $\sigma _0$ through $Z$ bosons. However, due to the LEP constraints, this interaction is largely suppressed by small mixing angles (of the order of $0.4\times 10^{-3}$). Although in the resonance $M_{\sigma}=M_Z/2$ the annihilation of DM enhance, it is not enough to reduce the relic density to the observable values. However, there arises a threshold where DM can annihilate to ordinary matter through interactions with the heavy gauge boson $Z'$, obtaining realistic relic density at $M_{\sigma}=70$ GeV.

On the other hand, by considering dispersions through scalar matter, we found smaller relic densities than through gauge interactions. For small Higgs couplings, we still obtain large relic densities for light DM. However, near the Higgs boson resonance $M_{\sigma}=M_{h_{0}}/2=63$ GeV, the annihilation rate sharply enhance to experimental limits on the relic density. If we extend the values of the parameters to wider ranges, we obtain larger allowed regions, with masses as low as $M_{\sigma} \approx 1.3 $ GeV, and as large as $M_{\sigma}\approx 125$ GeV. In particular, we obtain that the larger the scalar coupling constants, the larger the range for $M_{\sigma}$. 

The constraints on $M_{\sigma}$ and the thresholds exhibited by the relic denstiy may have observable consequences for DM research in experiments from direct detection and production in colliders, providing possible signatures in final states distributions that allow identify the presence of DM particles.

\section*{Appendix}

\appendix

\section{Z Pole observables\label{app:z-pole}}

The Z pole parameters with their experimental values from CERN
collider (LEP), SLAC Liner Collider (SLC) and data from atomic parity
violation taken from ref. \cite{data}, are shown in table \ref%
{tab:observables}, with the SM predictions and the expressions predicted by the extra
$U(1)_X$ model. The corresponding correlation matrix from ref. \cite{LEP} is
given in table \ref{tab:correlation}. We will use for the mass the approximation $M_{Z_1}=M_Z$.  In the SM, the partial decay widths of $Z_{1}$ into
fermion species $f_i\overline{f_{i}}$ is described by \cite{data, pitch}:

\begin{equation}
\Gamma _{i}^{SM}=\frac{N_{c}^{f}G_{f}M_{Z}^{3}}{6\sqrt{2}\pi }\rho _{i}%
\sqrt{1-\mu _{i}^{2}}\left[ \left( 1+\frac{\mu _{i}^{2}}{2}\right) \left(
v_{i}^{SM}\right) ^{2}+\left( 1-\mu _{i}^{2}\right) \left( a_{i}^{SM}\right)
^{2}\right] R_{QED}R_{QCD},  \label{partial-decay}
\end{equation}
where $N_{c}^{f}=1$, 3 for leptons and quarks, respectively. $%
R_{QED}=1+\delta _{QED}^{f}$ and $R_{QCD}=1+\frac{1}{2}\left(
N_{c}^{f}-1\right) \delta _{QCD}^{f}$ are QED and QCD corrections , and $\mu
_{i}^{2}=4m_{f_{i}}^{2}/M_{Z}^{2}$ considers kinematical corrections only
important for the $b$-quark. Universal electroweak corrections sensitive to
the top quark mass are taken into account in $\rho _{i}=1+\rho _{t}$ and in $%
v_{i}^{SM}$ which is written in terms of an effective Weinberg angle \cite%
{data}

\begin{equation}
\overline{S_{W}}^{2}=\left( 1+\frac{\rho _{t}}{T_{W}^{2}}\right) S_{W}^{2},
\label{effective-angle}
\end{equation}
with $\rho _{t}=3G_{f}m_{t}^{2}/8\sqrt{2}\pi ^{2}$. Nonuniversal
vertex corrections are also taken into account in the $Z_{1}\overline{b}b$
vertex with additional one-loop leading terms which leads to $\rho _{b}=1-%
\frac{1}{3}\rho _{t}$ and $\overline{S_{W}}^{2}=\left( 1+\frac{\rho _{t}}{%
T_{W}^{2}}+\frac{2\rho _{t}}{3}\right) S_{W}^{2}$ \cite{data, pitch}. 

For the top and bottom quark masses,
we use the following values calculated at the Z pole scale \cite{run-mass}:

\begin{eqnarray}
m_{t}(M_{Z}) &=&171.684 \  GeV, \nonumber \\
m_{b}(M_{Z})&=&2.853 \ GeV.  \label{quarks-mass}
\end{eqnarray}

For the partial SM partial decay given by Eq. (\ref{partial-decay}), we use
the following values taken from ref. \cite{data}

\begin{eqnarray}
\Gamma _{u}^{SM} &=&0.30026\pm 0.00005\text{ }GeV;\quad \Gamma
_{d}^{SM}=0.38304\pm 0.0005\text{ }GeV;  \notag \\
\Gamma _{b}^{SM} &=&0.37598\pm 0.00003\text{ }GeV;\quad \Gamma _{\nu
}^{SM}=0.16722\pm 0.00001\text{ }GeV;  \notag \\
\Gamma _{e}^{SM} &=&0.08400\pm 0.00001\text{ }GeV.  \label{SM-partial-decay}
\end{eqnarray}

For the $U(1)_X$ model in the fourth column of Tab. \ref{tab:observables}, we define from (\ref{modified-coup})  the coupling deviations 

\begin{eqnarray}
\delta{v_i^{SM}}&=&-\frac{g_XC_W}{g}v_i^{NSM}S_{\theta}, \ \ \ \ \  \delta{a_i^{SM}}=-\frac{g_XC_W}{g}a_i^{NSM}S_{\theta},
\end{eqnarray}
and we approximate $C_{\theta}=1$ for the mixing angle. Thus, the analytical expressions for the deviations of the Z pole observables are:

\begin{eqnarray}
\delta _{Z} &=&\frac{\Gamma _{u}^{SM}}{\Gamma _{Z}^{SM}}(\delta _{u}+\delta
_{c})+\frac{\Gamma _{d}^{SM}}{\Gamma _{Z}^{SM}}(\delta _{d}+\delta _{s})+%
\frac{\Gamma _{b}^{SM}}{\Gamma _{Z}^{SM}}\delta _{b}+3\frac{\Gamma _{\nu
}^{SM}}{\Gamma _{Z}^{SM}}\delta _{\nu }+3\frac{\Gamma _{e}^{SM}}{\Gamma
_{Z}^{SM}}\delta _{\ell };  \notag \\
\delta _{had} &=&R_{c}^{SM}(\delta _{u}+\delta _{c})+R_{b}^{SM}\delta _{b}+%
\frac{\Gamma _{d}^{SM}}{\Gamma _{had}^{SM}}(\delta _{d}+\delta _{s});  \notag
\\
\delta _{\sigma } &=&\delta _{had}+\delta _{\ell }-2\delta _{Z};  \notag \\
\delta A_{i} &=&\frac{\delta v_{i}^{SM}}{v_{i}^{SM}}+\frac{\delta 
a_{i}^{SM}}{a_{i}^{SM}}-\delta _{i},  \label{shift1}
\end{eqnarray}

\noindent where for the light fermions

\begin{equation}
\delta _{i}=\frac{2v_{i}^{SM}\delta v_{i}^{SM}+2a_{i}^{SM}\delta 
a_{i}^{SM}}{\left( v_{i}^{SM}\right) ^{2}+\left( a_{i}^{SM}\right)
^{2}},  \label{shift2}
\end{equation}

\noindent while for the $b$-quark

\begin{equation}
\delta _{b}=\frac{\left( 3-\beta _{K}^{2}\right) v_{b}^{SM}\delta v%
_{b}^{SM}+2\beta _{K}^{2}a_{b}^{SM}\delta a_{b}^{SM}}{\left( \frac{%
3-\beta _{K}^{2}}{2}\right) \left( v_{b}^{SM}\right) ^{2}+\beta
_{K}^{2}\left( a_{b}^{SM}\right) ^{2}}.  \label{shift3}
\end{equation}
where $\beta _K=\sqrt{1-(2m_b/M_Z)^2}$. The above expressions are evaluated in terms of the
effective Weinberg angle from Eq. (\ref{effective-angle}).

The weak charge is written as

\begin{equation}
Q_{W}=Q_{W}^{SM}+\Delta Q_{W}=Q_{W}^{SM}\left( 1+\delta Q_{W}\right) ,
\label{weak}
\end{equation}%
where $\delta Q_{W}=\frac{\Delta Q_{W}}{Q_{W}^{SM}}$. The deviation $\Delta
Q_{W}$ is \cite{cesio} 
\begin{equation}
\Delta Q_{W}=\left[ \left( 1+4\frac{S_{W}^{4}}{1-2S_{W}^{2}}\right) Z-N%
\right] \Delta \rho _{M}+\Delta Q_{W}^{\prime },  \label{dev}
\end{equation}%
and $\Delta Q_{W}^{\prime }$ which contains new physics gives

\begin{eqnarray}
\Delta Q_{W}^{\prime } &=&-16\left[ \left( 2Z+N\right) \left( a_{e}^{SM}%
{v}_{u}^{NSM}+{a}_{e}^{NSM}v_{u}^{SM}%
\right) +\left( Z+2N\right) \left( a_{e}^{SM}{v}%
_{d}^{NSM}+{a}_{e}^{NSM}v_{d}^{SM}\right) \right] S_{\theta } 
\notag \\
&&-16\left[ \left( 2Z+N\right) {a}_{e}^{NSM}{%
v}_{u}^{NSM}+\left( Z+2N\right) {a}_{e}^{NSM}{v}_{d}^{NSM}\right] \frac{M_{Z}^{2}}{M_{Z'}^{2}}.
\label{new}
\end{eqnarray}

For cesium, and for the first term in (\ref{dev}) we take the value $\left[ \left( 1+4%
\frac{S_{W}^{4}}{1-2S_{W}^{2}}\right) Z-N\right] \Delta \rho _{M}\simeq
-0.01 $ \cite{cesio2,cesio}.

\section*{Acknowledgment}

This work was supported by the Departamento Administrativo de Ciencia, Tecnolog\'{i}a e Innovaci\'on (COLCIENCIAS) in Colombia.

\newpage

\begin{table}[]
\begin{center}
\caption{\small Ordinary SM particle content, with $i=$1,2,3} \vspace{-0.5cm}
\label{tab:SM-espectro}
\begin{equation*}
\begin{tabular}{c c c c}
\hline\hline
$Spectrum$ & $G_{sm}$ & $U(1)_{X }$ & $Feature$\\ \hline \\
$\ 
\begin{tabular}{c}
$q^i_{L}=\left( 
\begin{array}{c}
U^i \\ 
D^i 
\end{array}%
\right) _{L}$ 
\end{tabular}%
\ $ & 
$(3,2,1/3)$
&
$\ 
\begin{tabular}{c}
$1/3$ for $i=3$ \\ 
$0$ for $i=1,2$%
\end{tabular}
\ $
&
chiral
\\ \\
$U_R^i$
& 
$(3^*,1,4/3)$
&
\begin{tabular}{c}
$2/3$ 
\end{tabular}
&
chiral
\\ \\
$D_R^i$
&
$(3^*,1,-2/3)$
&
\begin{tabular}{c}
$-1/3$ 
\end{tabular}
&
chiral
\\ \\
$\ 
\begin{tabular}{c}
$\ell ^i_{L}=\left( 
\begin{array}{c}
\nu ^i \\ 
e^i 
\end{array}%
\right) _{L}$ \\
\end{tabular}
\ $
& 
$(1,2,-1)$
&
$-1/3$ 
&
chiral
\\  \\
$e_R^i$
&
$(1,1,-2)$
&
\begin{tabular}{c}
$-1$ 
\end{tabular}
&
chiral
\\  \\
$\ 
\begin{tabular}{c}
$\phi _{1}=\left( 
\begin{array}{c}
\phi _1^{+} \\ 
\frac{1}{\sqrt{2}}(\upsilon _1+\xi _1+i\zeta _1) 
\end{array}%
\right) $
\end{tabular}
\ $
&
$(1,2,1)$
&
$2/3$
&
Scalar Doublet
\\  \\
$\ 
\begin{tabular}{c}
$W _{\mu }=\left( 
\begin{array}{cc}
W _{\mu }^{3}&\sqrt{2}W_{\mu }^+ \\ 
\sqrt{2}W_{\mu }^-& -W _{\mu }^{3}
\end{array}%
\right) $
\end{tabular}
\ $
&
$(1, 2 \times 2^*, 0)$
&
$0$
&
Vector
\\  \\
$\ 
B_{\mu }
\ $
&
$(1, 1, 0)$
&
$0$
&
Vector
\end{tabular}%
\end{equation*}%
\end{center}
\end{table}

\begin{table}[tbp]
\begin{center}
\caption{\small Exotic non-SM particle content, with $n=$1,2}\vspace{-0.5cm}
\label{tab:exotic-espectro}
\begin{equation*}
\begin{tabular}{c c c c}
\hline\hline
$Spectrum$ & $G_{sm}$ & $U(1)_{X }$ & $Feature$\\ \hline \\
$\ 
\begin{tabular}{c}
$T_L$ 
\end{tabular}%
\ $ & 
$(3,1,4/3)$
&
$1/3$
&
quasi-chiral
\\ \\
\begin{tabular}{c}
$T_R$
\end{tabular}
& 
$(3^*,1,4/3)$
&
$2/3$
&
quasi-chiral
\\ \\
\begin{tabular}{c}
$J_L^n$ 
\end{tabular}
&
$(3,1,-2/3)$
&
$0$
&
quasi-chiral
\\ \\ 
$\ 
\begin{tabular}{c}
$J_R^n$  
\end{tabular}
\ $
& 
$(3^*,1,-2/3)$
&
$-1/3$ 
&
quasi-chiral
\\ \\
\begin{tabular}{c}
$(\nu _R^i)^c$ 
\end{tabular}
&
$(1,1,0)$
&
$-1/3$
&
Majorana
\\ \\ 
\begin{tabular}{c}
$N _R^i$ 
\end{tabular}
&
$(1,1,0)$
&
$0$
&
Majorana
\\ \\
$\ 
\begin{tabular}{c}
$\phi _{2}=\left( 
\begin{array}{c}
\phi _2^{+} \\ 
\frac{1}{\sqrt{2}}(\upsilon _2+\xi _2+i\zeta _2) 
\end{array}%
\right) $ 
\end{tabular}
\ $
&
$(1,2,1)$
&
$1/3$
&
Scalar doublet
\\ \\
\begin{tabular}{c}
$\chi _0 = \frac{1}{\sqrt{2}}(\upsilon _{\chi}+\xi _{\chi }+i\zeta _{\chi})$ 
\end{tabular}
&
$(1,1,0)$
&
$-1/3$
&
Scalar singlet
\\ \\
\begin{tabular}{c}
$\sigma _0 = \frac{1}{\sqrt{2}}(\upsilon _{\sigma}+\xi _{\sigma}+i\zeta _{\sigma})$ 
\end{tabular}
&
$(1,1,0)$
&
$-1/3$
&
Scalar singlet
\\  \\ 
$Z'_{\mu}$ 
&
$(1,1,0)$
&
$0$
&
Vector
\end{tabular}%
\end{equation*}%
\end{center}
\end{table}

\begin{table}[]
\begin{center}
\caption{\small Vector and Axial couplings for the weak neutral currents $Z$ (SM-type) and $Z'$ (non-SM type) and for each fermion, with $i=1,2,3$, $a$ and $n=1,2$} \vspace{-0.5cm}
\label{tab:vector-axial-couplings}
\begin{equation*}
\begin{tabular}{c c c c c c}
 \hline \hline \vspace{0.1cm} 
$Fermion$&$v_i^{SM}$&$a_i^{SM}$&&$v_i^{NSM}$&$a_i^{NSM}$   \\  \hline \vspace{0.2cm}
$\nu ^i$ &$1/2$&$1/2$&&$1/3$&$1/3$ \\ \vspace{0.2cm}
$(\nu ^i)^c$&$0$&$0$&&$1/3$&$1/3$ \\ \vspace{0.2cm}
$N^i$&$0$&$0$&&$0$&$0$ \\ \vspace{0.2cm}
$e^i$&$-1/2+2S_W^2$&$-1/2$&&$4/3$&$-2/3$ \\ \vspace{0.2cm}
$U^3$&$1/2-4S_W^2/3$&$1/2$&&$-1$&$1/3$\\ \vspace{0.2cm}
$U^a$&$1/2-4S_W^2/3$&$1/2$&&$-2/3$&$2/3$\\ \vspace{0.2cm}
$D^3$&$-1/2+2S_W^2/3$&$-1/2$&&$0$&$-2/3$\\ \vspace{0.2cm}
$D^a$&$-1/2+2S_W^2/3$&$-1/2$&&$1/3$&$-1/3$\\ \vspace{0.2cm}
$T$ &$-4S_W^2/3$&$0$&&$-1$&$1/3$\\ \vspace{0.2cm}
$J^n$&$2S_W^2/3$&$0$&&$1/3$&$-1/3$
\end{tabular}
\end{equation*}
\end{center}
\end{table}

\begin{table}[h]
\caption{The parameters for experimental values, SM predictions and
$U(1)_X$ corrections. The values are taken from ref. \protect\cite{data}}
\begin{center}
$%
\begin{tabular}{cccc}
\hline
Quantity & Experimental Values & Standard Model & $U(1)_X$ Model \\ \hline\hline
$\Gamma _{Z}$ $\left[ GeV\right] $ & 2.4952 $\pm $ 0.0023 & 2.4961 $\pm $
0.0010 & $\Gamma _{Z}^{SM}\left( 1+\delta _{Z}\right) $ \\ \hline
$\Gamma _{had}$ $\left[ GeV\right] $ & 1.7444 $\pm $ 0.0020 & 1.7426 $\pm $
0.0010 & $\Gamma _{had}^{SM}\left( 1+\delta _{had}\right) $ \\ \hline
$\Gamma _{\left( \ell ^{+}\ell ^{-}\right) }$ $MeV$ & 83.984 $\pm $ 0.086 & 
84.005 $\pm $ 0.015 & $\Gamma _{\left( \ell ^{+}\ell ^{-}\right)
}^{SM}\left( 1+\delta _{\ell }\right) $ \\ \hline
$\sigma _{had}$ $\left[ nb\right] $ & 41.541 $\pm $ 0.037 & 41.477 $\pm $
0.009 & $\sigma _{had}^{SM}\left( 1+\delta _{\sigma }\right) $ \\ \hline
$R_{e}$ & 20.804 $\pm $ 0.050 & 20.744 $\pm $ 0.011 & $R_{e}^{SM}\left(
1+\delta _{had}+\delta _{e}\right) $ \\ \hline
$R_{\mu }$ & 20.785 $\pm $ 0.033 & 20.744 $\pm $ 0.011 & $R_{\mu
}^{SM}\left( 1+\delta _{had}+\delta _{\mu }\right) $ \\ \hline
$R_{\tau }$ & 20.764 $\pm $ 0.045 & 20.789 $\pm $ 0.011 & $R_{\tau
}^{SM}\left( 1+\delta _{had}+\delta _{\tau }\right) $ \\ \hline
$R_{b}$ & 0.21638 $\pm $ 0.00066 & 0.21576 $\pm $ 0.00004 & $%
R_{b}^{SM}\left( 1+\delta _{b}-\delta _{had}\right) $ \\ \hline
$R_{c}$ & 0.1720 $\pm $ 0.0030 & 0.17227 $\pm $ 0.00004 & $R_{c}^{SM}\left(
1+\delta _{c}-\delta _{had}\right) $ \\ \hline
$A_{e}$ & 0.15138 $\pm $ 0.00216 & 0.1475 $\pm $ 0.0010 & $A_{e}^{SM}\left(
1+\delta A_{e}\right) $ \\ \hline
$A_{\mu }$ & 0.142 $\pm $ 0.015 & 0.1475 $\pm $ 0.0010 & $A_{\mu
}^{SM}\left( 1+\delta A_{\mu }\right) $ \\ \hline
$A_{\tau }$ & 0.136 $\pm $ 0.015 & 0.1475 $\pm $ 0.0010 & $A_{\tau
}^{SM}\left( 1+\delta A_{\tau }\right) $ \\ \hline
$A_{b}$ & 0.925 $\pm $ 0.020 & 0.9348 $\pm $ 0.0001 & $A_{b}^{SM}\left(
1+\delta A_{b}\right) $ \\ \hline
$A_{c}$ & 0.670 $\pm $ 0.026 & 0.6680 $\pm $ 0.0004 & $A_{c}^{SM}\left(
1+\delta A_{c}\right) $ \\ \hline
$A_{s}$ & 0.895 $\pm $ 0.091 & 0.9357 $\pm $ 0.0001 & $A_{s}^{SM}\left(
1+\delta A_{s}\right) $ \\ \hline
$A_{FB}^{\left( 0,e\right) }$ & 0.0145 $\pm $ 0.0025 & 0.01633 $\pm $ 0.00021
& $A_{FB}^{(0,e)SM}\left( 1+2\delta A_{e}\right) $ \\ \hline
$A_{FB}^{\left( 0,\mu \right) }$ & 0.0169 $\pm $ 0.0013 & 0.01633 $\pm $
0.00021 & $A_{FB}^{(0,\mu )SM}\left( 1+\delta A_{e}+\delta A_{\mu }\right) $
\\ \hline
$A_{FB}^{\left( 0,\tau \right) }$ & 0.0188 $\pm $ 0.0017 & 0.01633 $\pm $
0.00021 & $A_{FB}^{(0,\tau )SM}\left( 1+\delta A_{e}+\delta A_{\tau }\right) 
$ \\ \hline
$A_{FB}^{\left( 0,b\right) }$ & 0.0997 $\pm $ 0.0016 & 0.1034 $\pm $ 0.0007
& $A_{FB}^{(0,b)SM}\left( 1+\delta A_{e}+\delta A_{b}\right) $ \\ \hline
$A_{FB}^{\left( 0,c\right) }$ & 0.0706 $\pm $ 0.0035 & 0.0739 $\pm $ 0.0005
& $A_{FB}^{(0,c)SM}\left( 1+\delta A_{e}+\delta A_{c}\right) $ \\ \hline
$A_{FB}^{\left( 0,s\right) }$ & 0.0976 $\pm $ 0.0114 & 0.1035 $\pm $ 0.0007
& $A_{FB}^{(0,s)SM}\left( 1+\delta A_{e}+\delta A_{s}\right) $ \\ \hline
$Q_{W}(Cs)$ & $-$73.20 $\pm $ 0.35 & $-$73.23 $\pm $ 0.02 & $%
Q_{W}^{SM}\left( 1+\delta Q_{W}\right) $ \\ \hline
\end{tabular}%
\ \ $%
\end{center}
\label{tab:observables}
\end{table}

\begin{table}[t]
\caption{The correlation coefficients for the Z-pole observables}
\begin{tabular}{ll}
\hline
$\Gamma _{had}$ & $\Gamma _{\ell }$ \\ \hline\hline
1 &  \\ 
.39 & 1 \\ \hline
\end{tabular}%
\par
\begin{tabular}{lll}
\hline
$A_{e}$ & $A_{\mu }$ & $A_{\tau }$ \\ \hline\hline
1 &  &  \\ 
.038 & 1 &  \\ 
.033 & .007 & 1 \\ \hline
\end{tabular}%
\par
\begin{tabular}{llllll}
\hline
$R_{b}$ & $R_{c}$ & $A_{b}$ & $A_{c}$ & $A_{FB}^{(0,b)}$ & $A_{FB}^{(0,c)}$
\\ \hline\hline
1 &  &  &  &  &  \\ 
-.18 & 1 &  &  &  &  \\ 
-.08 & .04 & 1 &  &  &  \\ 
.04 & -.06 & .11 & 1 &  &  \\ 
-.10 & .04 & .06 & .01 & 1 &  \\ 
.07 & -.06 & -.02 & .04 & .15 & 1 \\ \hline
\end{tabular}%
\par
\begin{tabular}{llllllll}
\hline
$\Gamma _{Z}$ & $\sigma _{had}$ & $R_{e}$ & $R_{\mu }$ & $R_{\tau }$ & $%
A_{FB}^{(0,e)}$ & $A_{FB}^{(0,\mu )}$ & $A_{FB}^{(0,\tau )}$ \\ \hline\hline
1 &  &  &  &  &  &  &  \\ 
-.297 & 1 &  &  &  &  &  &  \\ 
-.011 & .105 & 1 &  &  &  &  &  \\ 
.008 & .131 & .069 & 1 &  &  &  &  \\ 
.006 & .092 & .046 & .069 & 1 &  &  &  \\ 
.007 & .001 & -.371 & .001 & .003 & 1 &  &  \\ 
.002 & .003 & .020 & .012 & .001 & -.024 & 1 &  \\ 
.001 & .002 & .013 & -.003 & .009 & -.020 & .046 & 1 \\ \hline
\end{tabular}%
\label{tab:correlation}
\end{table}

\newpage


\end{document}